# A nonlinear plasma retroreflector for single pulse Compton backscattering

J.P. Palastro[1], D. Kaganovich[2], D. Gordon[2], B. Hafizi[2], M.Helle[2], J. Penano[2], and A. Ting[2]

[1]*Icarus Research, Inc., P.O. Box 30780, Bethesda, Maryland 20824-0780*
[2]*Plasma Physics Division, Naval Research Laboratory, Washington, DC 20375-5346*

**Abstract**

Compton scattered x-rays can be generated using a configuration consisting of a single, ultra-intense laser pulse, and a shaped gas target. The gas target incorporates a hydrodynamically formed density spike, which nonlinearly scatters the incident pump radiation, to produce a counter-propagating electromagnetic wiggler. This self-generated wiggler field Compton scatters from electrons accelerated in the laser wakefield of the pump radiation. The nonlinear scattering mechanism in the density spike is examined theoretically and numerically in order to optimize the Compton scattered radiation. It is found that narrow-band x-rays are produced by moderate intensity pump radiation incident on the quarter-critical surface of the density spike, while high fluence, broadband x-rays are produced by high intensity pump radiation reflected near the critical surface.

# I. Introduction

Relativistic electrons counterpropagating with respect to a laser pulse undergo rapid oscillatory motion in the pulse's electromagnetic field. This motion results in the emission of photons that travel predominately in the direction of the relativistic electrons and possess frequencies twice Doppler upshifted from the pulse's frequency. The emission, referred to as Compton backscattering [1], has a frequency $\omega_C \sim 4\gamma_0^2 \omega_0$, where $\omega_0$ is the pulse frequency and $\gamma_0$ the electron relativistic factor. Thus depending on the electron energy and pulse frequency, the emitted radiation can range from extreme ultraviolet, ~100 eV, to gamma, ~1 MeV, energies. With such a broad range of frequencies, Compton backscattering has been proposed for several applications including, phase contrast imaging [2], radiosurgery [3], lithography [4], and nuclear resonance fluorescence [5].

Radio-frequency linacs produce high-energy, low emittance electron beams ideal for Compton backscattering [6,7], but their size and cost are prohibitive for widespread deployment. All-optical Compton backscattering offers a potential small-scale alternative by replacing the linac with a laser wakefield accelerator. In a laser wakefield accelerator, the ponderomotive force of a high intensity, ultrashort laser pulse excites plasma waves whose fields can trap and accelerate electrons [8,9]. Standard all-optical Compton scattering therefore requires two ultrashort laser pulses, one for driving the wakefield to inject and accelerate electrons and one for scattering [10]. While both pulses can be generated from the same laser by splitting the initial pulse and aligning the resultant pulses for counterpropagation, the process can be cumbersome.

Ta Phuoc *et al.* demonstrated a single pulse scheme in which the wakefield driving pulse reflects from a self-generated plasma mirror at the surface of a CH or glass foil [11]. The reflected pulse then intersects the electrons trapped in the wakefield providing the wiggler. Solid foils have, however, limitations including erosion and bremsstrahlung emission from the accelerated electrons. Furthermore, the alignment issue is not completely eliminated: the pulse should intersect the foil at a small angle to ensure maximum overlap of the reflection and accelerated electrons.

Here we examine an alternative scheme for single ultrashort pulse Compton backscattering. A long (ns) pulse focused into a gas jet launches a shock wave. The shock wave and gas jet flow collide forming a sharp density spike [12,13]. The leading edge of an incident ultrashort laser pulse ionizes the gas. The bulk of the pulse undergoes a partial Poynting flux reversal from the ionized spike, providing a counterpropagating field. This field Compton backscatters from electrons accelerated in the wakefield driven by the incident pulse. A schematic of the interaction is displayed in Fig. (1).

The major advantages of this scheme lie in the high rep-rate of the gas jet system and, as we will see, the significant retro-reflection of light from the spike. Optimizing the Compton radiation requires characterization of the counterpropagating field and an understanding of the reversal mechanism. Using the full format PIC simulation TurboWAVE [14], we investigate the Poynting flux reversal and resulting Compton spectrum as a function of incident pulse amplitude and plasma spike profile.

For normal incidence, we simply refer to the reversal as a reflection. In this case we find that there are two regimes depending on the density of the spike. For low spike densities $n_{sp} < n_{cr}$, where $n_{cr} = m_e \omega_0^2 / 4\pi e^2$ is the critical density, the reflection occurs predominately at the quarter critical surface, resulting in a backscattered spectrum centered near $\omega_0/2$. This reflection is a general phenomenon occurring when high intensity light interacts with a quarter critical surface [15-18]. Previous studies, however, were motivated by laser-based inertial confinement fusion and thus characterized by low intensity, $I < 10^{17}$ W/cm$^2$, long pulses, $\tau > 10$ ps, and hot plasmas, $T_e > 500$ eV. More recent observations of these reflections have been made in the context of highly oblique angle ultrashort pulse-solid interactions for electron heating [19,20]. Our interest here is in high intensity $I > 5 \times 10^{17}$ W/cm$^2$, ultrashort pulses, $\tau \leq 50$ fs in cold plasmas. Here we make the connection between this reflection and the saturation of the absolutely unstable Raman instability [15], which by itself would only allow stationary scattering. Bonnaud *et al.* suggested a similar mechanism for long pulses, in which the absolute Raman instability seeds the backscattered convective wave [18].

Most of the incident pulse energy makes it past the quarter critical layer. For higher spike densities $n_{sp} \sim n_{cr}$, a second, more efficient reflection occurs as the result of a stationary, periodic electron density fluctuation that develops near the critical surface. The ponderomotive force generated by the spatial overlap of the reflected and incident pulse drive the electron density fluctuation. The periodic density fluctuation modifies the local plasma dielectric constant, providing a Bragg-like grating that enhances the reflection. The resulting reflected spectrum is centered near $\omega_0$.

We also show that the properties of the Compton X-ray spectrum are determined primarily by the amplitude of the incident pulse and are relatively insensitive to the spike density. For instance, if the incident pulse amplitude is low, $a_0 = 0.5$, the reflected pulse is narrow band, providing a narrow band spectrum characteristic of small amplitude Compton scattering [21]. If, on the other hand, the incident pulse amplitude is large, $a_0 > 1.0$, the reflected pulse is nonlinear, providing a broadband Compton X-ray spectrum [21]. Finally, we find that for non-normal incidence of the pulse, a significant fraction of energy is reflected directly backward, ensuring the robustness of our scheme to small alignment errors. Specifically, the insensitivity of the Compton X-ray spectrum is demonstrated for incidence angles between 1° and 5°.

The remainder of the manuscript is organized as follows. Section II details the set up for our simulations. In Section III we present simulation results for pulses normally incident on the density spike and explain the mechanisms behind the nonlinear reflection. This includes a discussion of 3D effects. In Section IV we present the Compton spectra resulting from normally incident pulses and discuss the effects leading to the spectral features. Section V includes our results for the nonlinear reflection and Compton scattering from pulses non-normally incident on the density spike. Section VI concludes the paper with a summary of our results.

**II. Simulation Setup**

To investigate the laser pulse-density spike interaction we conducted particle-in-cell (PIC) simulations using TurboWAVE [14]. TurboWAVE is a framework for solving the Maxwell-Lorentz system of equations for charged particle dynamics. The model is fully relativistic and fully electromagnetic. The fields are advanced with a Yee solver [22], and the sources are deposited using quadratic weighting, with charge conservation ensured by means of a Poisson solver.

The fields, particle trajectories, densities, and currents were calculated on a 2D or 3D planar-Cartesian grid in the lab frame. The 3D simulation domain was 183 μm × 183 μm × 214 μm in the transverse, $x$ and $y$, and longitudinal directions, $z$, respectively. For the 2D simulations the $y$ direction was eliminated. The cell sizes were chosen to resolve the laser wavelength along the propagation direction. The specific values will be presented with each case. The relatively short axial domain served two purposes. First, because our interest lies in the properties of the reflected light, the simulations must be conducted in the lab frame as opposed to the moving frame. This places a higher computational cost on the total simulated distance, which can be mitigated by considering a shorter axial domain. Second, the shorter axial domain ensures that the properties of the incident pulse are well characterized because there isn't enough path length for nonlinear propagation to occur before the spike.

The plasma density profile was chosen to represent the intersection of a hydrodynamic shock launched in a heavy gas with a low density, light gas region. In the simulation, the gas was pre-ionized, and the ions were immobile providing only a neutralizing background. The electron density initially ramped up over 24 μm to a density of $2\times10^{19}$ cm$^{-3}$. Following a plateau of 50 μm, the electron density again ramped up over 10 μm to peak density, which was varied between $2\times10^{20}$ cm$^{-3}$ and $1.8\times10^{21}$ cm$^{-3}$. After the peak, the density returned, over 20 μm, to a value of $2\times10^{20}$ cm$^{-3}$ where it remained until a final down ramp of 24 μm. The simulated density profiles are illustrated in Fig. (2).

The laser pulse was initialized with x or y linear polarization, a sin$^2$ temporal profile, and a Gaussian transverse profile. The pulse parameters were as follows: central wavelength λ = 800 nm, temporal full width half maximum (FWHM) τ = 50 fs, and an exp(−1) field width w = 23 μm. The normalized field amplitude, $a_0 = eE/m_e c\omega_0$, where $\omega_0 = 2\pi c/\lambda$, $m_e$ is the electron mass, $c$ the speed of light, and $e$ the fundamental unit of charge, was varied between 0.5 and 2.0. These parameters correspond to pulse energies between 0.22 and 3.5 J. The pulse started with its front edge at the beginning of the density ramp and centered along the z-axis. To change the laser pulse's angle of incidence with respect to the spike, the spike profile was rotated in the x-z plane, while keeping the propagation axis aligned with the z-axis.

**III. Normal incidence: nonlinear reflection**

We begin by considering the reflection of normally incident p-polarized laser pulses from spikes of varying density. We conducted a single 3D simulation to verify the phenomena observed in 2D simulations, which we present later in this section. The 3D simulation domain was discretized with 256 × 256 × 8192 cells in the transverse, $x$ and $y$, and longitudinal directions, $z$, respectively. Figure (3) illustrates the 3D transmission and reflection of an $a_0 = 1.0$ pulse from a spike of density $1 \times 10^{21}$ cm$^{-3}$ or 0.57 $n_{cr}$ where $n_{cr} = m_e \omega_0^2 / 4\pi e^2 = 1.7 \times 10^{21}$ cm$^{-3}$ is the critical density. The figure displays snapshots of the pulse intensity envelope before during and after the reflection at t = 280 fs, t = 340 fs, t = 400 fs, and t = 460 fs. In each frame, the incident or transmitted pulse propagates from the lower right to the upper left. At 400 fs the reflected and transmitted pulses have spatially separated. The reflected pulse emanates from the center of the incident pulse resulting in the centralized intensity depletion in the transmitted pulse by 460 fs. Surprisingly, the reflected pulse's peak intensity is comparable to that of the incident pulse. The local depletion and reflection demonstrate nonlinearity: the reflected and transmitted energy depend on the local intensity of the incident pulse. The major difference between the 3D and 2D results was in the intensity of the reflected pulse. Nonlinear focusing in the high-density plasma led to a 1.5x enhancement in the peak reflected intensity in 3D. Other than the intensity enhancement, the spatial profiles in both 2D and 3D were quite similar.

We continue by considering 2D simulations, which allow investigation of a wider range of parameters. This is particularly useful for optimizing the reflected pulse for Compton backscattering. For the following results, the simulation domain was discretized with 256 × 16384 cells. Once the transmitted and reflected pulses become spatially distinct, we can calculate the fraction of energy reflected and transmitted by integrating over the region of space containing each pulse. In particular we define the reflection fraction, $R$, and transmission fraction, $T$, as

$$R = \frac{1}{4\pi U_L} \int d^2 \mathbf{x}_\perp \int_0^{z_S - \Delta} \mathbf{E}^2 \, dz \quad \text{(1a)}$$

$$T = \frac{1}{4\pi U_L} \int d^2\mathbf{x}_\perp \int_{z_S+\Delta}^{z_{max}} \mathbf{E}^2 \, dz \quad \text{(1b)}$$

where $z_S$ marks the peak of the spike, $\Delta$ is the width of the spike, $z_{max}$ is the axial simulation boundary, and $U_L$ is the incident laser pulse energy. The use of $\Delta$ eliminates electrostatic energy contributions within the spike, which along with the plasma kinetic energy account for the absorbed pulse energy. In Eq. (1) we have approximated the total electromagnetic energy density by twice the contribution from the electric field. We have verified that the electrostatic contribution to $R$ and $T$ outside of the spike is small in regions devoid of the pulse (less than 1% of the energy density localized in the spike). From Eq. (1) we can also obtain the absorption fraction in the interaction: $A = 1 - R - T$.

Figure (4) displays the reflection and absorption fractions as a function of peak spike density. The general trends are that $R$ increases with electron density, and $A$ increases with $a_0$. From $a_0 = 1.0$ to $a_0 = 2.0$, the increase in $A$ coincides with the drop in $R$. We note that, in general, the peak reflected amplitude, $a_R$, is greater than $R^{1/2}a_0$. Several effects contribute to the sharp increase in reflected and absorption fractions for $a_0 > 0.5$. Foremost, we have already seen in Fig. (3) that the reflection is nonlinear. The area of the pulse that can drive the nonlinearity and, as a result, be depleted increases with $a_0$. To understand the other contributing effects and nonlinearity in more detail, we turn to Fig. (5). Figure (5) shows the spatial power spectral density (PSD) of the reflected pulse for each incident $a_0$ and $n_{sp}$. The red dashed and solid lines mark k = $k_0$ and k = $k_0/2$ respectively, and each PSD curve has been normalized to its maximum. The $n_e = 2 \times 10^{20}$ cm$^{-3}$ curve for $a_0 = 0.5$ has been omitted because the reflected energy was nearly zero. For each incident pulse amplitude, the PSD exhibits two qualitative behaviors depending on the spike density. At low spike densities, $4 \times 10^{20}$ cm$^{-3}$ < $n_{sp}$ ≤ $1.0 \times 10^{21}$ cm$^{-3}$, the spectral peak occurs near $k_0/2$, while at higher densities, $n_{sp} \geq 1.4 \times 10^{21}$ cm$^{-3}$, the peak is shifted towards k = $k_0$.

At the lower spike densities the reflection results from the saturation of the absolute Raman backscattering instability. Near the quarter critical density, $n_{cr}/4 \sim 4 \times 10^{20}$ cm$^{-3}$, the incident electromagnetic pulse undergoes resonant Raman scattering and decays into a scattered electromagnetic and electrostatic wave each with a frequency $\omega_s \sim \omega_0/2 \sim$ $\omega_p$. The scattering is absolutely unstable and both scattered waves are trapped in a region near the quarter critical surface where they undergo temporal growth [15]. Drake and Lee have analyzed the instability in detail [15]. In their analysis, the parameter $N = 2^{-1/2} a_0^{3/2} k_0 L$ quantifies the strength of the inhomogeneity, where $L$ is the gradient scale length at the quarter critical surface, $L \sim [\partial_z \ln(n_e)]|_{z=z_{qc}}$, and $z_{qc}$ is the axial position of the surface. Values of $N \gg 1$ and $N \ll 1$ indicate weak and strong imhomogeneity respectively. The lower spike densities, $n_{sp} < 1.4 \times 10^{21}$ cm$^{-3}$, have scale lengths $L > 2.4$ $\mu$m, and thus $N > 5$ for each $a_0$ we consider. Even though the gradients are comparable to the laser wavelength, the relatively large amplitudes place our parameters well within the weakly inhomogeneous regime.

For weak inhomogeneity, the temporal growth rate, $\Gamma$, scattered frequency of the electromagnetic wave, $\omega_{EM}$, and total width of the absolutely unstable region, $\delta z$, are given by the following:

$$\frac{\Gamma}{\omega_{pq}} \cong \frac{\sqrt{3}}{2} a_0 \left[ 1 - \left(\frac{1}{12}\right)^{1/4} N^{-1} \right] \quad (2a)$$

$$\frac{\omega_{EM}}{\omega_{pq}} \cong \left[ 1 + 4\sqrt{3} \left(\frac{\omega_{pq}}{\Gamma}\right)^{1/2} N^{-5} \right] \quad (2b)$$

$$k_{pq} \delta z = \left(\frac{3}{2}\right)^{1/2} a_0^{-1/2} N \left(\frac{k_{pq}}{k_0}\right) \left[ 1 - \frac{4}{3} a_0^{-2} \left(\frac{\Gamma}{\omega_{pq}}\right)^2 \right]^{1/2} \quad (2c)$$

where $\omega_{pq}^2 = \pi e^2 n_{cr} / m_e$ is the plasma frequency at the quarter critical surface. To estimate the effect of relativistic pump amplitudes, we can apply to Eq. (2) the transformations $\omega_{pq}^2 \to \gamma_\perp \omega_{pq}^2$ and $a_0 \to a_0 / \gamma_\perp^{1/2}$ where $\gamma_\perp = (1 + a_0^2/2)^{1/2}$ is the relativistic factor accounting the electron's rapid motion in the pump field. Figure (4) indicates a density dependence of the reflected and absorbed energy fraction, but $\Gamma$ and $\omega_{EM}$ in Eq. (2) only depend weakly on density through $N$ by way of $L$. So where then does the density dependence of Fig. (4) originate? The local group velocity of the incident pulse can be written $v_{g0}(z)/c \sim 1 - \omega_p^2(z)/2\omega_0^2$. In a sharp, increasing density gradient, the group velocity at the back of the pulse is larger than that in the front. This results in a compression of the pulse or "swelling" that increases the pulse amplitude. Assuming fixed energy and a flat-top temporal pulse shape, one can show that the amplitude increases as $a_{sw}(z) \sim a_0 \exp[(\omega_{pq}^2/4\omega_0^2)(z/L)]$. Taking $z$ = 5 μm we find $a_{sw}/a_0 \sim$ 1.07, 1.11, and 1.14 for $n_{sp}$ = 6.0, 8.0, and 10.0 $\times 10^{20}$ cm$^{-3}$ respectively. These small changes in the pulse amplitude increase $\exp(\Gamma\tau)$ substantially.

Equation (2) and the swelling model illustrate three points. First, the growth rate of the instability and hence the reflected and absorbed energy increase with both $a_0$ and $n_{sp}$ consistent with Fig. (4). As an example, using a spike density $n_{sp}$ = 1.0×10$^{21}$ cm$^{-3}$, the incident pulse FWHM as the exponentiation time, and the simple estimate of compression, the gains are $\Gamma\tau \sim$ 22, 50, and 100 for $a_0$ = 0.5, 1.0, and 2.0 respectively. Second, the frequency of the scattered electromagnetic wave is nearly $\omega_{pq}$ consistent with Fig. (5). Finally, the absolute instability is localized to a region of ~ 7 μm. The absolute Raman theory also explains the reduced downshifting of the spectral peak for a spike density of $n_{sp}$ = 2×10$^{20}$ cm$^{-3}$ < $n_{cr}$/4 and amplitudes $a_0 \geq$ 1.0 in Fig. (5).

The analysis of Drake and Lee [15] provides a framework to understand the reflection, but their treatment is limited to a non-evolving continuous, plane pump wave and the linear stage of instability. For examining non-ideal effects, we return now to the simulations. Figure (6) displays contributions to the electromagnetic energy density from different spectral ranges and field components as a function of time and axial coordinate. The incident pulse amplitude and spike density are $a_0$ = 0.5 and $n_{sp}$ = 1×10$^{21}$ cm$^{-3}$ respectively. The contributions have been averaged over the transverse coordinate and normalized to their peak values for clarity. The dashed white line marks the quarter critical surface. In Fig. (6a) the energy density of the incident pulse is plotted. The contours have a positive slope representing forward propagation. The drop off in energy density occurs due to the reduction in wavenumber as the plasma density increases: for fixed frequency $ck_z \sim [\omega^2 - \omega_p^2(z) - c^2 k_\perp^2]^{1/2}$. In Fig. (6b) we see the electrostatic energy density initially grow in time and remain stationary (non-convecting) at the quarter critical surface. Figure (6c) shows the stationary, near zero wavenumber, localized growth, of the electromagnetic energy at the quarter critical surface. These first three subplots illustrate the expected energetics of the three waves interacting in the absolute Raman instability.

We have yet to explain why an absolute instability leads to the emission of a convecting electromagnetic wave, aka our reflected pulse. In both Fig. (6b) and (6c) we see the energy density grow rapidly in time and then abruptly crash (the electrostatic energy density goes through several such cycles). The emission of the reflected pulse occurs simultaneously with this crash as can be seen in Fig. (6d) where the reflected pulse appears as the negative slope contours. Thus we conclude that the reflected pulse results from the absolute Raman instability's saturation. The seeding of the convective wave by the absolute Raman instability has also been suggested in the context of low-intensity, long pulse-quarter critical interactions [18]. One could also put forth the explanation that the trapping "potential" disappears as the incident pulse convects away from the quarter critical region. However, Fig. (6) shows that the incident pulse is still within this region.

The dynamics are similar for larger amplitudes, $a_0$ = 1.0 to $a_0$ = 2.0, but the growth, saturation, and emission cycles can occur several times. Nonlinear evolution of the incident wave, including depletion and spectral broadening, introduces additional complexity. In spite of this, a simple explanation can be offered for the broad spectra redshifted beyond $k_0/2$ observed in Fig. (5). The reflected pulse originates in a high-density region possessing a wavelength twice that of the pump and an amplitude nearly as large. Each of these, the high density, long wavelength, and high amplitude, make the reflected pulse itself susceptible to Raman instabilities. For instance, as the reflected pulse propagates into lower densities, it encounters its own quarter critical surface where it undergoes absolutely unstable Raman scattering, while at slightly lower densities it undergoes Raman forward scattering [23,24]. This additional nonlinear scattering leads to the complicated spectra at lower spike densities, $4\times10^{20}$ cm$^{-3}$ < $n_{sp}$ ≤ $1.0\times10^{21}$ cm$^{-3}$, for $a_0$ = 1.0 and $a_0$ = 2.0.

The incident pulse retains much of its energy after interacting with the initial critical surface: more than 95% for the parameters considered in Fig. (6). For lower spike densities, $n_{sp}$ ≤ $1.4\times10^{21}$ cm$^{-3}$, what remains of the incident pulse interacts with the quarter critical surface at the back of the spike before propagating away. For larger spike densities, the leading edge of the pulse undergoes a partial reflection from the near critical density surface. The ponderomotive force generated by the spatial overlap of the pulse and its reflection sets up a stationary periodic density fluctuation: $\delta n \sim \cos(k_0 z - \omega_0 t)\cos(k_0 z + \omega_0 t)$ or $\delta n \sim \cos(2k_0 z)$. This density fluctuation, $\delta n$, when added to the background electron density can exceed the critical density, $n_0 + \delta n > n_{cr}$, where $n_0$ is the local electron density, while the periodicity provides a Bragg-like grating. An ideal Bragg grating consists of alternating layers of materials with different dielectric constants. The light incident on each interface contributes a Fresnel (linear) reflection that can add constructively with the reflections from all other interfaces, leading to a reflection far surpassing that of a single dielectric slab. Here the oscillations in electron density act as the alternating layers.

Figure (7) displays the on-axis intensity, calculated from $E_x^2$, (right scale, blue) and absolute electron density fluctuation (left scale, red) enveloped at the incident pulse wavenumber as a function of axial coordinate at 337 fs. The incident pulse amplitude and peak spike density are $a_0$ = 1.0 and $n_{sp}$ = 1.4×10$^{21}$ cm$^{-3}$ respectively. The left dashed black line marks the quarter critical surface, and the right dashed black line the location of the peak spike density. The figure demonstrates the large periodic electron density fluctuation set up by the ponderomotive force of the incident and reflected light. As seen in Fig. (4), the reflection fraction is relatively large for $a_0$ = 1.0 and 2.0 when $n_{sp}$ =1.4×10$^{21}$ cm$^{-3}$ in spite of the background electron density being lower than the critical density. This is a direct result of the large density fluctuation acting as a Bragg-like mirror. Additionally, the weak reflection of $a_0$ = 0.5 when $n_{sp}$ =1.4×10$^{21}$ cm$^{-3}$ results from the pulse amplitude being too small to set up a large density fluctuation.

As a comparison, we consider the reflection fraction from a single dielectric slab predicted by Fresnel theory:

$$R_F = \frac{\eta_L^2 + \eta_R^2 + 2\eta_L \eta_R \cos(2k_S \Delta)}{1 + \eta_L^2 \eta_R^2 + 2\eta_L \eta_R \cos(2k_S \Delta)} \quad (3)$$

where $\eta_L = (\varepsilon_S^{1/2} - \varepsilon_L^{1/2})/(\varepsilon_S^{1/2} + \varepsilon_L^{1/2})$, $\eta_R = (\varepsilon_R^{1/2} - \varepsilon_S^{1/2})/(\varepsilon_R^{1/2} + \varepsilon_S^{1/2})$, $\varepsilon_x = 1 - \omega_{px}^2/\omega_0^2$ is the dielectric function of the plasma, and $L$, $S$, and $R$ denote the regions left of the spike, the spike itself, and right of the spike. For $n_{sp}$ = 1.4×10$^{21}$ cm$^{-3}$, upon taking $\Delta$ = 9 μm and setting $k_S = \varepsilon_S^{1/2} \omega_S / c$, Eq. (3) predicts $R_F = 0.07$. This is comparable to the value calculated by Eq. (1a) for $a_0 = 0.5$, $R = 0.03$, but much smaller than the value for $a_0 = 1.0$, $R = 0.35$. The disagreement cannot be explained by including the effect of relativistic transparency through the transformation $\omega_p^2 \to \omega_p^2/\gamma_\perp$ in Eq. (3). In fact, this transformation makes the disagreement with $a_0 = 1.0$ even worse. Furthermore, Eq. (3) is sensitive to $\Delta$: it predicts small reflection fractions over limited ranges of $\Delta$, while the reflections we observe are quite robust.

However, as discussed above, the $a_0 = 1.0$ pulse drives a periodic density fluctuation resulting in a total electron density, $n_{sp} + \delta n$, sufficient to set up a Bragg mirror that enhances the reflection. For an ideal Bragg grating the reflection fraction is given by

$$R_B = \left| \frac{\kappa \sinh(\Gamma \ell)}{\Gamma \sinh(\Gamma \ell) - i\delta \sinh(\Gamma \ell)} \right|^2 \quad (4)$$

where $\Gamma = (\kappa^2 - \delta^2)^{1/2}$, $\kappa = (k_0/\pi)(\varepsilon_{max}^{1/2} - \varepsilon_{min}^{1/2})$, $\delta = <\varepsilon^{1/2}> k_0 - \pi/\lambda_B$, $<\varepsilon^{1/2}>$ is the average refractive index in the grating, $\lambda_B$ is the periodicity of the grating, and $\ell$ is the total length of the grating [25]. Using Fig (7) as an example, we set $\delta n = 1 \times 10^{20}$ cm$^{-3}$, $\lambda_B$ = 800 nm, and $\ell$ = 4 µm and find $R_B$ = 0.42. This reflection fraction is comparable but greater than that observed in Fig. (4). A small disagreement, however, is not surprising: the ideal analysis neglects the changes in $\delta n$ along the spike, the transmission of light through the back of the spike, and absorption. The Bragg-mirror picture also explains the limited reflection for $a_0 = 0.5$. The ponderomotive force when $a_0 = 0.5$ is too weak to generate a $\delta n$ large enough for reflection.

In support of a Bragg-like reflection, we observe smaller shifts in the wavenumber of the reflected pulse than in the absolute Raman mechanism. This can be observed in Fig. (5). When $a_0 = 0.5$ and $a_0 = 1.0$ with $n_{sp} = 1.4 \times 10^{21}$ cm$^{-3}$, spectral signatures are present at $k_0/2$ and $k_0$ corresponding to the reflections from the quarter critical and near-critical density surfaces respectively. Again the complicated spectral structure for $a_0 = 1.0$ and $a_0 = 2.0$ can be explained by additional Raman scattering of the reflected wave as it propagates out of the spike region.

## IV. Normal incidence: Compton Scattering

In our scheme, the reflected pulse Compton backscatters from electrons accelerated in the wakefield driven by the incident pulse. The bandwidth, amplitude, and direction of the reflected pulse will determine the bandwidth and energy fluence of the Compton x-rays. For instance, a narrow band, low amplitude scattering pulse will result in a narrow band x-ray spectrum with low fluence, while a broadband, high amplitude scattering pulse will result in a broadband spectrum with high fluence. The connection between the pulse and x-ray spectrum is made through the trajectories of the relativistic electrons. Esarey *et al.* derived expressions for the trajectories of relativistic electrons counterpropagating with respect to an optical field. In their treatment, the optical pulse depends only on the moving frame coordinate $\xi = z + ct$, and finite pulse duration and diffraction effects are ignored [21,26]. As seen in Fig. (3) it is not clear that these assumptions are valid for our reflected pulses. We will refer to the trajectories and resulting Compton spectra of Esarey *et al.* as the ideal case.

In particular, upon adapting their results for our reflected pulse, we have

$$x = x_0 + \frac{a_R}{(1+\beta_0)\gamma_0 k_R}\sin(k_R \xi) \quad (5a)$$

$$z = z_0 + \frac{1}{2}\left[1 - \frac{\gamma_\perp^2}{\gamma_0^2(1+\beta_0)^2}\right]\xi - \frac{a_R^2}{8\gamma_0^2(1+\beta_0)^2 k_R}\sin(2k_R\xi) \quad (5b)$$

where $a_R$ and $k_R$ are the amplitude and wavenumber of the reflected light respectively, $\gamma_0 = (1-\beta_0^2)^{-1/2}$ and, $\beta_0 = v_{z0}/c$ is the initial normalized electron velocity. With the analytic trajectories, the radiated spectrum can be calculated from

$$\frac{d^2U}{d\omega d\Omega} = \frac{e^2\omega^2}{4\pi^2 c}\left|\int dt\left[\mathbf{n}\times(\mathbf{n}\times\boldsymbol{\beta})\right]\exp\left[i\omega(t - \mathbf{n}\cdot\mathbf{x}/c)\right]\right|^2 \quad (6)$$

where $\Omega$ is the solid angle, $\boldsymbol{\beta} = \mathbf{v}/c$ is the normalized electron velocity, and **n** is the unit vector pointing to the location of observation [27]. Esarey *et al.* evaluate Eq. (6) in the limit $a_R \ll 1$ and $a_R \gg 1$. In the low amplitude limit, Eq. (6) evaluated in the forward direction is given by

$$\left.\frac{d^2U}{d\omega d\Omega}\right|_{\theta=0} = \frac{e^2}{8\pi c}a_R^2(k_R L_{int})\left(\frac{\omega_C}{\omega_R}\right)\omega_C \delta(\omega-\omega_C), \quad (7a)$$

while in the high amplitude limit

$$\left.\frac{d^2U}{d\omega d\Omega}\right|_{\theta=0} = \frac{6}{\pi^3}\frac{e^2}{c}(k_R L_{int})\gamma_0^2\left(\frac{\omega}{\omega_K}\right)^2 K_{2/3}^2\left(\frac{\omega}{\omega_K}\right), \quad (7b)$$

where $L_{int}$ is the interaction distance of the reflected pulse and relativistic electrons, half the reflected pulse duration for instance, $\omega_K = 3a_R^3\gamma_0^2/(1+a_R^2/2)$, and $K$ is the modified Bessel function. Equation (7) shows that, in the ideal case, a low amplitude pulse results in a narrow x-ray spectrum centered about $\omega_C = 4a_R^2\gamma_0^2$, while a high amplitude pulse results in a broad spectrum with characteristic width $\omega_K$. This is precisely what we observe in the simulations.

Figure (8) displays single particle Compton radiated spectra in the forward, positive z direction, resulting from reflections of incident pulses with $a_0 = 0.5$, $a_0 = 1.0$, and $a_0 = 2.0$ and spike densities $n_{sp} = 1.0\times10^{21}$ cm$^{-3}$ and $n_{sp} = 1.8\times10^{21}$ cm$^{-3}$. The frequency is normalized to $4\gamma_0^2\omega_0$. The spectra were calculated from Eq. (6) using trajectories of test particles with $\gamma_0 = 19.6$ started on-axis, 22 µm behind the peak of the laser pulse and evolved in the self-consistent fields of the simulation. Spectra were also calculated for initial energies of $\gamma_0 = 98.0$, 196, 391, and 978, but are not shown because the spectral features were nearly identical. We note that due to the short simulation distance, the test electrons were not accelerated in the pulse's wakefield and were meant to be representative of wakefield accelerated electrons. The densities were chosen such that scattering pulses were representative of reflections from the quarter critical surface and the critical surface. In each case, the peak amplitude of the reflected pulse was comparable to that of the incident pulse.

The $a_0 = 0.5$ pulse results in a narrow band spectrum for both densities with peaks at $\omega/4\gamma_0^2\omega_R \sim 1$, where $\omega_R = ck_R$: $\omega_R = \omega_0/2$ or $\omega_0$ for $n_{sp} = 1.0\times10^{21}$ cm$^{-3}$ or $n_{sp} = 1.8\times10^{21}$ cm$^{-3}$ respectively. Interestingly, at $n_{sp} = 1.8\times10^{21}$ cm$^{-3}$ the Compton spectrum exhibits a signature of the pulse reflection at $n_c/4$ as a tail extending from $\omega=1$ to $1/2$. For both $a_0 = 1.0$ and $a_0 = 2.0$, the gross spectrum is broad but is not a smooth curve as predicted by Eq. (7b). This results from not being in the asymptotic limit of $a_R \gg 1.0$, the finite temporal limits of integration in Eq. (6), and finite pulse duration effects such as the electron interacting with a range of amplitudes. The $a_0 = 1.0$ and $a_0 = 2.0$ pulse result in more total energy: when $n_{sp} = 1.8 \times 10^{21}$ cm$^{-3}$, $dU/d\Omega = 26$ and 40 keV compared to 6 keV for $a_0 = 0.5$.

The incoherence of the reflected field can also result in non-ideal electron motion, which may also modify the spectral properties of the Compton radiation. The degree of spatial coherence for the transverse component of the reflected field, $E_W = \hat{\mathbf{x}} \cdot \mathbf{E}_R$ (wiggler field) evaluated on-axis is given by

$$G(z) = \frac{\int E_W(z') E_W(z'-z) dz'}{\int E_W^2(z') dz'} \quad , (8)$$

with the associated coherence length $\ell_z = \int |G(z)|^2 dz$. For $a_0 = 0.5$, 1.0, and 2.0 respectively we find $\ell_z = 16$, 15, and 12 μm respectively. While the wiggler field coherence length is similar for $a_0 = 0.5$ and 1.0, the Compton spectrum for $a_0 = 0.5$ is narrow while that of $a_0 = 1.0$ is broad. We then conclude that the broad Compton spectrum is, in fact, the result of large amplitude 'ideal' Compton scattering, and not the lack of wiggler coherence.

As an example of a possible experimental realization, we consider a weakly nonlinear pulse, $a_0 = 0.5$, that has injected 1 nC of electrons via a density down ramp [28]. The electrons are accelerated in the quasi-linear wake of the pulse up to 10 MeV. The reflection of the pulse from the hydrodynamically formed density spike then Compton scatters from the electrons resulting in a sum of on-axis single particle spectra observed in Fig. (8). Using the trajectories from our simulations, we integrate Eq. (6) numerically over frequency and solid angle to find a total energy into X-ray of 300 GeV and a total photon number of $2 \times 10^8$. We note that the total energy into X-rays goes up as $\gamma_0^2$ while the photon number remains essentially constant.

To summarize, the reflections of low amplitude pulses from the quarter critical surface are ideal for narrowband x-ray spectra, while reflections of high amplitude pulses from the critical surface are ideal for high fluence, broadband x-rays.

**V. Non-normal incidence**

For practical use, our single pulse Compton scattering scheme must be insensitive to alignment errors. If, for instance, the incident pulse encounters the density spike at an oblique angle, and undergoes a specular reflection, the reflected light could miss the electron beam entirely. To examine the alignment sensitivity, we performed a series of simulations in which the incidence angle of the laser pulse onto the spike was varied. Four angles were considered: $\theta$ = 0°, 1°, 5°, and 22.5°. The spatial grid for each angle had to resolve the laser wavelength in the incident and specularly reflected directions. This required increasing the spatial resolution in the transverse direction. Specifically, 256, 512, 2048, and 8192 transverse cells were used for $\theta$ = 0°, 1°, 5°, and 22.5° respectively. In all cases the pulse was p-polarized. S-polarization will be considered in future studies.

Here we refer to the pulse travelling in the negative z-direction as the scattered pulse to avoid ambiguity in the term "reflection," and we refer to specular and retro-reflections as they apply. Figures (9) and (10) display the spatial power spectral densities (PSDs) of the scattered light for each angle of incidence on a $\log_{10}$ scale. Here the pulse amplitudes are $a_0 = 0.5$ and $a_0 = 1.0$ in Figs. (9) and (10) respectively, and $n_{sp} = 1.8 \times 10^{21}$ cm$^{-3}$. The inner and outer white lines mark circles of wavenumber $|\mathbf{k}| = k_0/2$ and $k_0$ respectively. The horizontal and diagonal dashed lines trace the retro and specularly reflected directions respectively.

Figure (9) shows that for $a_0 = 0.5$ the scattered pulse propagates primarily in the specular direction with wavenumber $|\mathbf{k}| = k_0$. We thus conclude that the specular reflection for all angles at $a_0 = 0.5$ is dominated by a Fresnel-like reflection at the critical surface. Recall from Sec. III that the amplitude is too small to result in the ponderomotive mechanism. Small signatures from the quarter critical layer interaction can be observed at $|\mathbf{k}| = k_0/2$. Although dominated by the specular reflection, a significant fraction of the pulse energy is retro-reflected at small angles, $\theta = 1°- 5°$, which we return to below.

Figure (10) shows that for $a_0 = 1.0$ the scattered pulse propagates at a range of angles varying from the retro to specularly reflected directions with wavenumbers $|\mathbf{k}| \sim k_0$. In Sec. III, we found that a normal incident reflection can result in a stationary ponderomotive electron fluctuation that results in a Bragg-like mirror. For non-normal incidence, a similar fluctuation develops from specular reflections: $\delta n \sim \cos(k_z z + k_x x - \omega_0 t)\cos(k_z z - k_x x + \omega_0 t)$ or $\delta n \sim \cos(2k_0 z)$. If a retro-reflection is seeded, a fluctuation develops of the form $\delta n \sim \cos(k_z z + k_x x - \omega t)\cos(k_z z + k_x x - \omega t)$ or $\delta n \sim \cos[2(k_z z + k_x x)]$, which is constant along a line $x = -(k_z / k_x)z$ and thus enhances the retro reflection through the Bragg-like mechanism.

The enhancement of spectral signal at $|\mathbf{k}| \sim k_0/2$ for $\theta = 22.5°$ suggests that the absolute Raman scattering is more efficient at larger angles of incidence. Afeyan and Williams [29] derived the oblique angle correction, $g$, for the absolute Raman growth:

$$\frac{\Gamma}{\omega_{pq}} \cong a_0 g(\theta, k_x) \left[1 - \left(\frac{3}{16}\right)^{1/2} \frac{1}{Ng^{3/2}}\right] \quad (9a)$$

$$g(\theta, k_x) = \left[\varepsilon \cos^2\theta + \left(\frac{k_x}{k_0} - \varepsilon^{1/2} \sin\theta\right)^2\right]^{1/2}, \quad (9b)$$

where $k_x$ is the wavenumber of the scattered wave. As an example, the growth rate for the specularly reflected light is nearly 1.5x larger for $\theta = 22.5°$ than it is for $\theta = 0°$. Furthermore, with non-normal incidence the pulse can refract from the transverse density gradient, increasing the angle of incidence before the interaction with the scattering surface.

Figure (11) displays the Compton spectral energy density in the forward, positive z, direction as a function of scaled frequency, $4\gamma_0^2 \omega_0$, for, $\theta = 0°$, 1°, and 5°, and $a_0 = 0.5$ (bold, blue curve) and $a_0 = 1.0$ (narrow, red curve). The Compton spectrum is relatively insensitive to the incidence angle: the spectral features are similar for all three angles, and we even observe an enhancement in the spectral amplitude for $\theta = 1°$. Because the reflected pulse is wider than it is long, a small angle increases the interaction length, which, in turn, enhances the spectral amplitude. We note that part of the insensitivity to angle can be explained geometrically. The relativistic electrons start 22 μm behind the peak, thus if the scattering takes place at the peak of the pulse, the specular reflection would need a width smaller than 4 μm to miss the electrons entirely.

**VI. Summary and Conclusions**

We have investigated, through simulation, a novel scheme for single pulse Compton backscattering. The schemes relies on the intersection of a hydrodynamically launched shock wave and gas jet flow to form an underdense to near critical density spike [12,13]. The spike, once ionized by the leading edge of a pulse, acts as a plasma retroreflector. As a result, one pulse can both drive a wakefield to accelerate a population of electrons and backscatter from these electrons. Simulations in the weakly nonlinear regime predict that the Compton scattering of the reflected beam from 1 nC of 10 MeV electrons results in ~$10^8$ 1.5 keV photons.

Depending on the spike density and pulse amplitude, one can tune the reflection for the desired Compton spectrum. For small incident pulse amplitudes, $a_0 = 0.5$, the Compton spectrum is narrow band with lower total energy, while for larger incident pulse amplitudes the Compton spectrum is broadband with higher total energy. One can imagine more complicated hydrodynamic shock geometries, tailoring the pulse shape or phase [30], employing pulse trains, or using several electron bunches to further tune the spectrum, ideas we are currently pursuing. As an example, the pulse could be shaped with a long, low amplitude pedestal, which, upon reflection, would increase the interaction length of the Compton scattering and ensure a narrow X-ray spectrum [31].

**Acknowledgements**

This work is supported by the Department of Energy and the Naval Research Laboratory 6.1 Program. The authors would like to thank Y.-H. Chen, P. Sprangle, and S. Kalmykov for fruitful discussions.

**Figures**

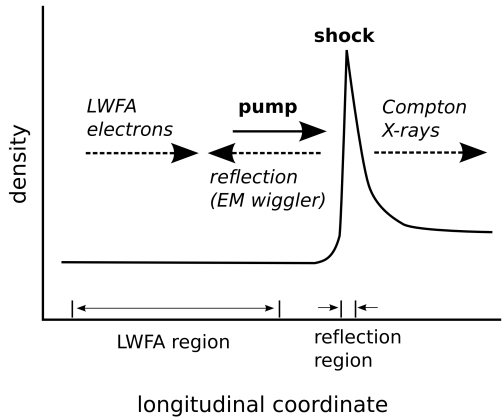

Figure 1. Schematic of the single pulse Compton scattering scheme. The bold text and solid lines represent externally provided components, while the italic text and dashed lines are self-generated components.

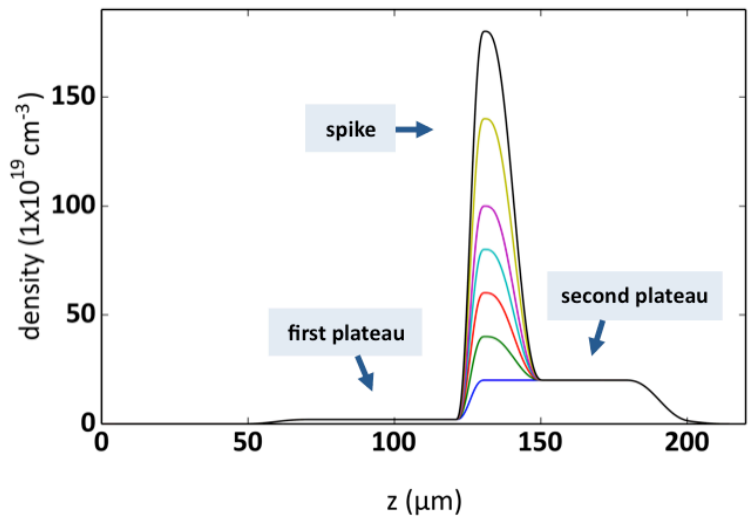

Figure 2. Simulated plasma density profiles representing the intersection of a hydrodynamic shock launched in a heavy gas with a low density, light gas. The incident laser pulse propagates from left to right.

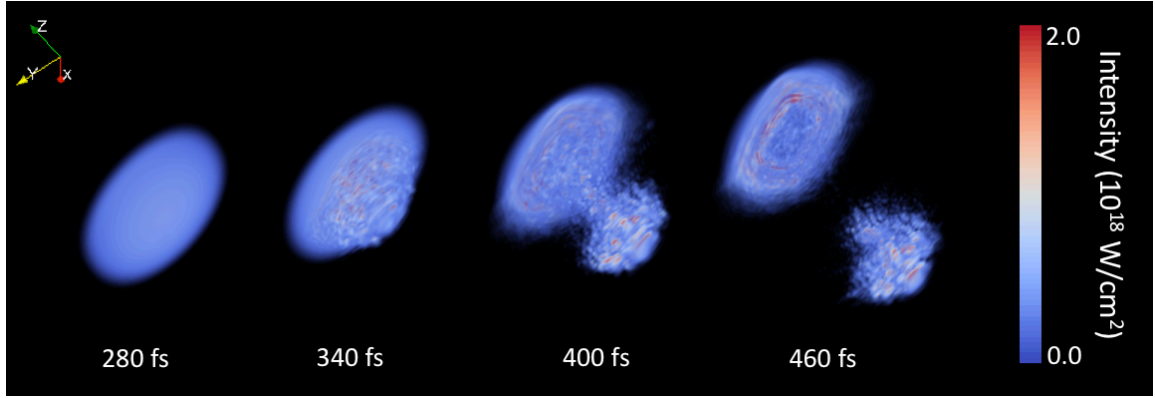

Figure 3. Sequences of frames illustrating the 3D reflection and transmission of an $a_0 = 1.0$ pulse from a plasma spike with peak electron density $1 \times 10^{21}$ cm$^{-3}$. The colors represent isocontours of intensity and, in each frame, the incident pulse is propagating from the lower right to the upper left.

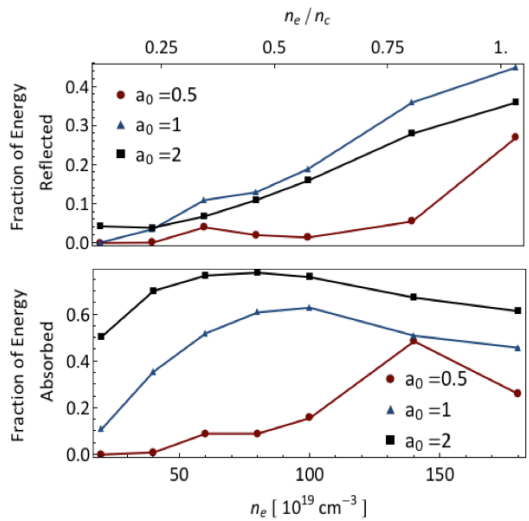

Figure 4. Fraction of energy reflected (top) and absorbed (bottom) as a function of peak spike electron density for three different incident pulse amplitudes, $a_0 = 0.5$, $a_0 = 1.0$, and $a_0 = 2.0$.

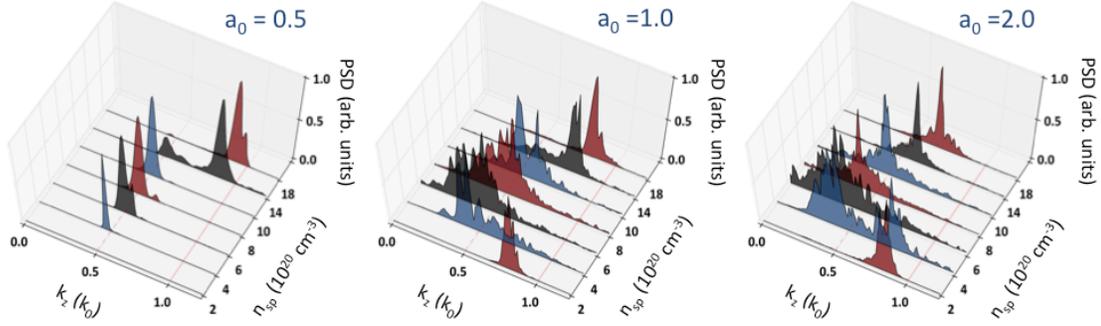

Figure 5. Power spectral density (PSD) of the reflected pulse normalized to the maximum for several values of peak spike electron density and three different incident pulse amplitudes, $a_0 = 0.5$, $a_0 = 1.0$, and $a_0 = 2.0$. The red solid and dashed lines mark $k = k_0$ and $k = k_0/2$ respectively. The $n_e = 2 \times 10^{20}$ cm$^{-3}$ curve for $a_0 = 0.5$ has been omitted because the reflected energy was nearly zero.

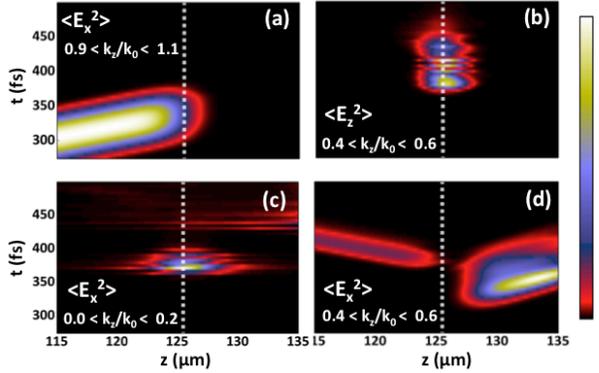

Figure 6. Contributions to the electromagnetic and electrostatic energy density from different spectral ranges and field components as a function of time and axial coordinate. The incident pulse amplitude and spike density are $a_0 = 0.5$ and $n_{sp} = 1 \times 10^{21}$ cm$^{-3}$ respectively. The contributions have been averaged over the transverse coordinate and normalized to their peak values. The dashed white line marks the quarter critical surface. (a) energy density of the incident pulse, (b) electrostatic energy density, (c) near zero wavenumber electromagnetic energy density, (d) energy density of the reflected pulse and transmitted light.

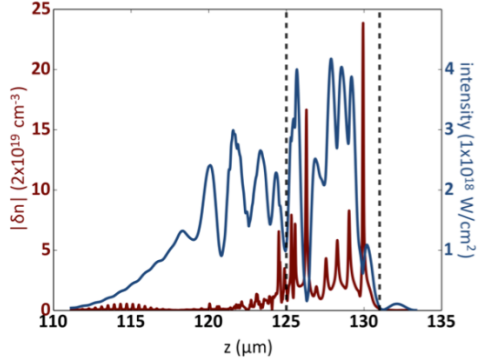

Figure 7. Intensity (right scale, blue) and absolute electron density fluctuation (left scale, red) enveloped at the incident pulse wavenumber as a function of axial coordinate after 337 fs. The incident pulse amplitude and peak spike density are $a_0 = 1.0$ and $n_{sp} = 1.4\times10^{21}$ cm$^{-3}$ respectively. The left dashed black line marks the quarter critical surface, and the right dashed black line marks the location of the peak spike density.

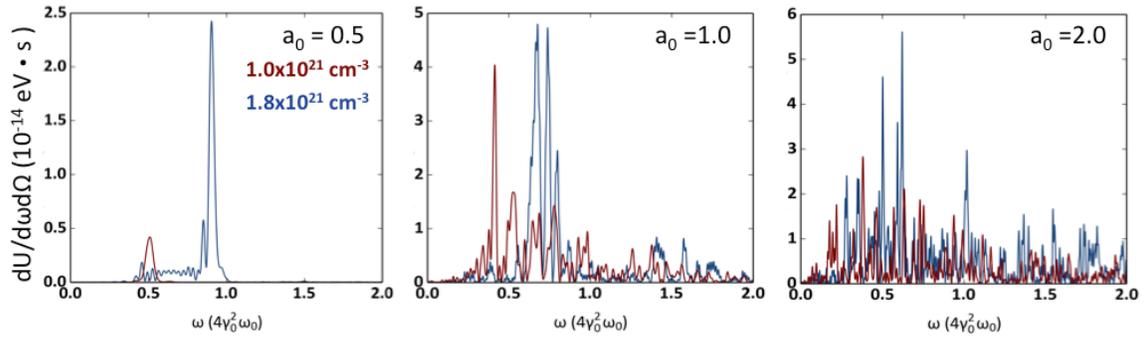

Figure 8. Compton scattered spectral energy density in the forward, positive z, direction as a function of scaled frequency, $4\gamma_0^2\omega_0$, for three different incident pulse amplitudes, $a_0 = 0.5$, $a_0 = 1.0$, and $a_0 = 2.0$, and two different densities, $n_e = 1.0\times10^{21}$ cm$^{-3}$ and $n_{sp} = 1.8\times10^{21}$ cm$^{-3}$

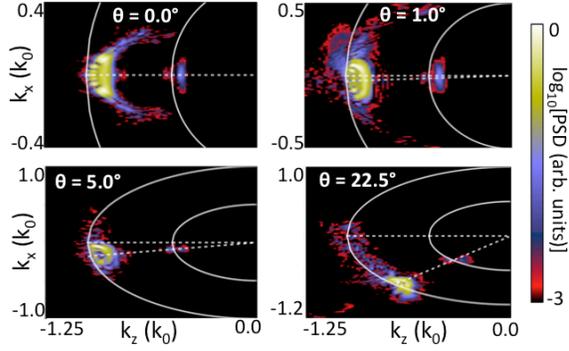

Figure 9. Spatial spectrum of the scattered light for three different incidence angles of a pulse with $a_0 = 0.5$. The peak spike density was $n_{sp} = 1.8 \times 10^{21}$ cm$^{-3}$. The inner and outer solid white lines mark circles of wavenumber $|\mathbf{k}| = k_0/2$ and $k_0$ respectively. The horizontal and diagonal dashed lines trace the retro and specularly reflected directions respectively.

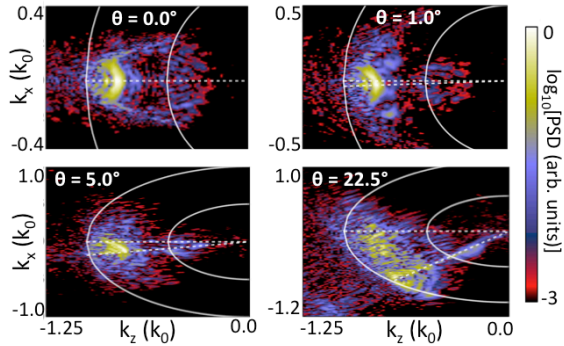

Figure 10. Spatial spectrum of the scattered light for three different incidence angles of a pulse with $a_0 = 1.0$. The peak spike density was $n_{sp} = 1.8 \times 10^{21}$ cm$^{-3}$. The inner and outer solid white lines mark circles of wavenumber $|\mathbf{k}| = k_0/2$ and $k_0$ respectively. The horizontal and diagonal dashed lines trace the retro and specularly reflected directions respectively.

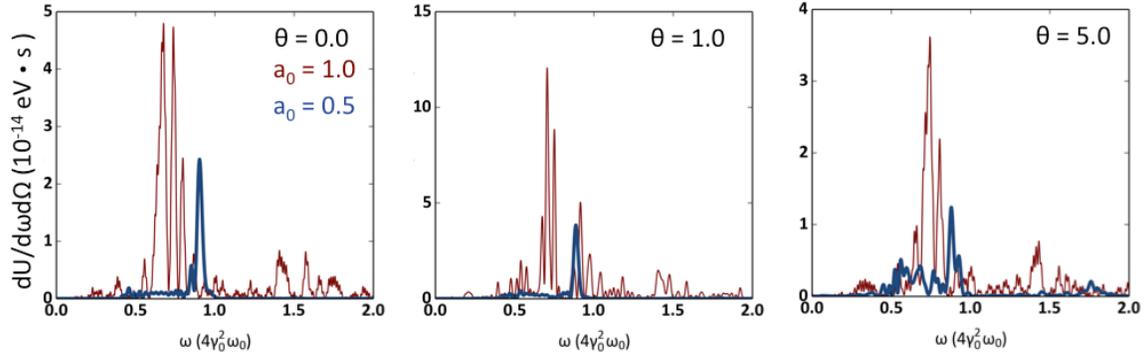

Figure 11. Compton scattered spectral energy density in the forward, positive z, direction as a function of scaled frequency, $4\gamma_0^2\omega_0$, for three different incident angles, $\theta = 0°$, $1°$, and $5°$, and two incident amplitudes $a_0 = 0.5$ (bold, blue curve) and $a_0 = 1.0$ (red, narrow curve).